\documentclass[lettersize,journal]{IEEEtran}
\usepackage[utf8]{inputenc}
\usepackage[T1]{fontenc}

\usepackage{enumitem}
\usepackage{amsmath, amssymb}
\usepackage{semantic}
\usepackage{nicematrix}
\usepackage[caption=false,font=normalsize,labelfont=sf,textfont=sf]{subfig}
\usepackage{arydshln}
\usepackage{algpseudocode}
\usepackage{algorithm}
\usepackage{graphicx}
\usepackage{comment}
\usepackage{booktabs}
\usepackage{pifont}
\newcommand{\xmark}{\ding{55}}%
\usepackage{balance}
\usepackage{multirow}
\usepackage{optidef}
\usepackage{mathrsfs}
\usepackage{tablefootnote}
\usepackage{hyperref}
\usepackage{cite}
\usepackage{makecell}
\usepackage{varwidth}
\usepackage{tikz,xcolor,hyperref}
\definecolor{lime}{HTML}{A6CE39}
\DeclareRobustCommand{\orcidicon}{
	\begin{tikzpicture}
	\draw[lime, fill=lime] (0,0) 
	circle [radius=0.16] 
	node[white] {{\fontfamily{qag}\selectfont \tiny ID}};
	\draw[white, fill=white] (-0.0625,0.095) 
	circle [radius=0.007];
	\end{tikzpicture}
	\hspace{-2mm}
}
\foreach \x in {A, ..., Z}{\expandafter\xdef\csname orcid\x\endcsname{\noexpand\href{https://orcid.org/\csname orcidauthor\x\endcsname}
			{\noexpand\orcidicon}}
}
\usepackage[compact]{titlesec}         
\titlespacing{\section}{8pt}{8pt}{8pt} 
\AtBeginDocument{
  \setlength\abovedisplayskip{2pt}
  \setlength\belowdisplayskip{2pt}}

\newtheorem{definition}{Definition}

\newtheorem{assumption}{Assumption}

\newcommand{\Rn}{\mathbb{R}^n}
\newcommand{\Rnn}{\mathbb{R}^{n \times n}}

\newcommand{\cac}[1]{\textcolor{cyan}{#1}}
\newcommand{\mac}[1]{\textcolor{magenta}{#1}}
\usepackage{float}
\usepackage{soul}
\allowdisplaybreaks
\def\BibTeX{{\rm B\kern-.05em{\sc i\kern-.025em b}\kern-.08em
    T\kern-.1667em\lower.7ex\hbox{E}\kern-.125emX}}
\begin{document}

\title{Effects of Line Dynamics on Stability Margin to Hopf Bifurcation in Grid-Forming Inverters}
\author{Sushobhan Chatterjee\orcidA{} and Sijia Geng\orcidB{}
\thanks{The authors are with the Department of Electrical and Computer Engineering, Johns Hopkins University. \texttt{(email: schatt21@jhu.edu; sgeng@jhu.edu)}.}%
}

\maketitle

\begin{abstract}
This paper studies the parameter sensitivity of grid-forming inverters to Hopf bifurcations to address oscillatory instability. An analytical expression for the sensitivity of the stability margin is derived based on the normal vector to the bifurcation hypersurface. We identify the most effective control parameters through comprehensive analysis. In particular, the impacts of line dynamics on the stability margin to Hopf bifurcation are investigated.
The results indicate that the feedforward gain in the voltage control loop is the most effective parameter for enhancing the stability margin. Furthermore, it is observed that line dynamics introduce a uniform reduction in the stability margin across all parameters. However, the reduction is generally small for most parameters except for the voltage-reactive power droop gain, which shows a more pronounced effect.
\end{abstract}


\section{Introduction} \label{intro}
As inverter-based resources (IBRs) are becoming more prevalent, maintaining stability has become a significant challenge \cite{cheng2022real,modi2024replication}. 
A \textit{grid-forming} (GFM) inverter functions as a controlled voltage source with flexibility in power output, leveraging storage or curtailment to actively support grid stability. The most common GFM control strategies emphasize frequency-power droop control \cite{chandorkar1993control}. Other prominent methods involve emulating the physics and control mechanisms of synchronous machines \cite{zhong2010synchronverters}, or designing inverters to act like virtual oscillators \cite{torres2015synchronization,sinha2015uncovering}, etc.

When studying the stability of conventional power systems, the dynamics of transmission lines are often ignored (i.e., assumed to be of zeroth order and thereby modeled by algebraic power flow equations or voltage-current relationships as preserved by the network admittance matrix). This approach is valid in conventional power systems that are dominated by synchronous machines with slow response times (order of seconds) as supported by multi-timescale approximations \cite{peponides1982singular}. However, as IBRs are capable of operating at much faster time scales (order of milliseconds), neglecting transmission line dynamics in the model may impact system stability assessment. 
This issue has been observed in droop-controlled microgrids \cite{mariani2014model} and is experimentally evident for the previously mentioned control schemes.
\cite{vorobev2017high} analyzes this problem and derives intuitive bounds on the droop gains that are obtained using small-signal analysis for reduced-order GFM inverters, while \cite{geng2022unified} studies the impact of static and dynamic lines on the transient performance of a unified IBR controller. \cite{gross2019effect} uses singular perturbation and Lyapunov functions to analyze the impacts of dynamic lines on dVOC inverter's performance. 

The scope of designing corrective actions or control tunings to mitigate such instabilities is not well studied in the literature. Given an instability condition, it was shown in \cite{chatterjee2024sensitivity} that a ``proper tuning'' of system and control parameters can successfully restore a stable system equilibrium.
Ensuring system stability can be formulated as a geometric problem within a multidimensional parameter space \cite{dobson1992observations}. This approach involves determining the position of nominal parameters, $\lambda_0$, relative to hypersurfaces where stability is lost (i.e., bifurcations). The proximity of $\lambda_0$ to the nearest bifurcations defines the stability margin. The sensitivity of the stability margin to parameters helps in determining the optimal parameter adjustments for improving stability robustness \cite{dobson2003distance}. 
An analytical approach for examining parameter sensitivity to bifurcations uses the normal vector method, as described by \cite{dobson1992iterative}. Building on this, \cite{dobson1993new} and \cite{dobson1992sensitivity} investigated oscillatory instabilities in traditional power systems. The impact of parameters on various bifurcations in grid-following (GFL) IBRs was qualitatively analyzed by \cite{yang2022bifurcations} and quantitatively in our previous work \cite{chatterjee2024sensitivity}. 

In this paper, we focus on Hopf bifurcation in GFM inverters with detailed modeling and control structures and systematically analyze the parameter sensitivity to oscillatory instabilities. Specifically, we account for transmission line dynamics and its impacts on stability compared to static line models. Using normal vector theory, we derive analytical expressions for calculating parameter sensitivities and stability margins of a full-order GFM inverter-based system to Hopf bifurcation.

\section{Problem Formulation} \label{prob}
Our main goal is to examine the impacts of parameters and line dynamics on the stability boundaries of GFM inverters. To this end, we summarize the {normal vector} method in this section and present an analytical formula for the parameter sensitivity of the stability margin concerning the Hopf bifurcations. For a more complete derivation, refer to \cite{chatterjee2024sensitivity}.

\subsection{Notations}
Let $\Rn$ and $\mathbb{C}^n$ be the space of $n$-dimensional real and complex vectors, respectively, and $\Rnn$ be the space of real square matrices of order $n$. Let $\mathbb{S}^n$ be a hypersphere of unit length in $\mathbb{R}^n$. Let $I_n$ be the $n \times n$ identity matrix. For a matrix $A \in \mathbb{R}^{n \times n}$, $\mu_i(A)$ denotes its $i$th eigenvalue. For a vector $x \in \mathbb{C}^n$, $x_i$ denotes its $i${th} element, $\bar{x}$ denotes its complex conjugate, $x^T$ and $x^H$ denotes its transpose and conjugate transpose (i.e. $\bar{x}^T$), respectively, and $\Vert x \rVert$ denotes its 2-norm unless stated otherwise. For a scalar $x$, $|x|$ denotes its absolute value. $\Re(x)$ denotes the real part of a complex number. For function $f(x,y)$, $D_xf$ denotes the partial derivative of $f$ with respect to $x$.

\subsection{Hopf Bifurcation}
Consider a system described by differential equations:
\begin{equation} \label{eqn:e1}
    \dot{x} = f(x, \lambda),
\end{equation}
where $f: \mathbb{R}^{n} \times \mathbb{R}^{m} \rightarrow \mathbb{R}^{n}$ is a nonlinear and smooth function, $x \in \mathbb{R}^n$ represents the state variables, and $\lambda \in \mathbb{R}^m$ denotes the parameters. For a parameter vector $\lambda_0 \in \mathbb{R}^m$, which indicates the system's nominal or current parameters, we define an equilibrium of equation \eqref{eqn:e1} as $x_0$ and assume it is asymptotically stable. As $\lambda$ changes within the parameter space $\mathbb{R}^m$, the equilibrium point $x_0$ shifts within the state space $\mathbb{R}^n$ and may vanish or become unstable due to a bifurcation. 
In general, oscillations arising in power systems are usually accompanied by strange nonlinear attractors born from the onset of Hopf bifurcations \cite{varghese1998bifurcations}. The formal definition of the Hopf bifurcation is given below.
\begin{definition} \label{defn:d1}
    \textit{\cite[Hopf Bifurcation]{seydel2009practical} Assume that $f$ is a twice continuously differentiable function, and,
    \begin{enumerate}
        \item $f(x_{*},\lambda_{*}) = 0$,
        \item $f_x(x_{*},\lambda_{*})$ possess a simple pair of purely imaginary eigenvalues $\mu(\lambda_{*}) = \pm j \omega_{*},\ \omega_{*} > 0$, and does not have any other eigenvalues with zero real part,
        \item $D_{\lambda}(\Re \{\mu(\lambda)\})|_{*} \neq 0$.
    \end{enumerate}
    Then, limit cycles bifurcate from the steady-state solution at the equilibrium point $(x_{*},\lambda_{*})$. This phenomenon is termed as Hopf bifurcation.} 
\end{definition}

In the ensuing analysis, let \( x_{*} \) and \( \lambda_{*} \) represent the equilibrium and parameter values, respectively, at a bifurcation, and let \( f_x|_{*} = f_x|_{(x_{*}, \lambda_{*})} \) denote the Jacobian \( f_x \) evaluated at that bifurcation. The Hopf bifurcation hypersurfaces, denoted by \( \Sigma^H \subseteq \mathbb{R}^m \), correspond to the set of \( \lambda_{*} \) values for which equation \eqref{eqn:e1} undergoes a Hopf bifurcation at \( (x_{*}, \lambda_{*}) \). Designing measures to prevent the system from approaching the bifurcation boundaries is crucial for intercepting oscillations. 

\subsection{Normal Vector to Bifurcation Hypersurface}
This section derives and interprets the formula for computing normal vectors to the hypersurfaces of Hopf bifurcations in parameter space.
Let $j\omega_{*}$ be an eigenvalue of $f_{x}|_{(x_{*}, \lambda_{*})}$. By definition, since $f_x|_{*}$ is invertible, the Implicit Function Theorem guarantees the existence of a smooth function $u$ and a scalar constant $\delta > 0$, defined in the neighborhood of $z_{*}:= (x_{*}, \lambda_{*})$, such that $x = u(\lambda)$ and $f(u(\lambda), \lambda) = 0$ for all $\lVert z - z_{*} \rVert \leq \delta$. 
Under appropriate non-degeneracy and transversality conditions
(see \cite[Sec 3.4]{guckenheimer2013nonlinear} for further details), $\Sigma^H$ has a normal vector $\mathcal{N}(\lambda_{*}) \in \mathbb{R}^m$ at $\lambda_{*} \in \Sigma^H$,
and the corresponding Gauss map $\mathcal{N}: \Sigma^H \rightarrow \mathbb{S}^m$ is smooth.
The expression for the normal vector for Hopf bifurcations can be derived as \cite{chatterjee2024sensitivity}\footnote{The normal vector formula described by \eqref{eqn:e5} embeds a natural geometric interpretation of the transversality conditions $D_{\lambda}(\Re\{\mu(\lambda)\})|_{*} \neq 0$.},
\begin{align} \label{eqn:e5}
    \mathcal{N}&(\lambda_{*}) := \beta \Re\Big\{w^H\big(-f_{xx} f_x^{-1} f_{\lambda} + f_{x \lambda}\big)v\Big\}\Big|_{*},
\end{align} 
where $\beta \neq 0$ is a real scaling factor, and $v \in \mathbb{C}^n$ and $w \in \mathbb{C}^n$ are the normalized right and left eigenvectors, respectively, such that $\lVert v \rVert = 1$ and $w^Hv = 1$.

\subsection{Parameter Sensitivity of the Stability Margins}
The stability margin, defined as the distance to the nearest bifurcation from a nominal parameter $\lambda_0$, is given by $\Delta(\lambda_0) = \lVert \lambda_{*} - \lambda_0 \rVert$. Adjusting $\lambda_0$ to increase $\Delta$ is desirable especially when $\Delta$ is small. The optimal direction for the first-order changes in $\lambda_0$ to maximize $\Delta$ is determined by the sensitivity $\Delta_{\lambda_0} = \frac{\partial \Delta(\lambda_0)}{\partial \lambda_0}$. In other words, we treat the margin $\Delta$ as a function of the parameters $\lambda$ and seek the direction of largest sensitivity $\Delta_{\lambda_0} \in \mathbb{R}^m$ to enlarge the margin by adjusting parameters in that direction.

\begin{figure}[ht!]
    \leftline{\includegraphics[scale=1.4]{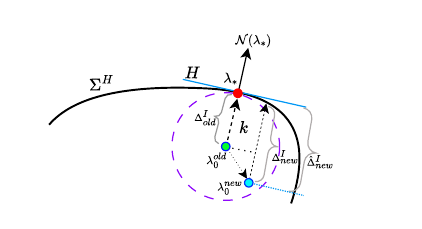}}
    \caption{Geometry of the bifurcation hypersurface, normal vectors, and the stability margin.}
    \label{fig:norm_vec} 
\end{figure}

Consider a parameter variation from $\lambda_0$ along a direction $k \in \mathbb{R}^m$. The sensitivity $\Delta_{\lambda_0}$ represents a scaled projection of the normal vector in the direction of $k$, as illustrated in Fig. \ref{fig:norm_vec}.
Using the condition of detecting the first bifurcation as parameters vary in the direction $k$,
\begin{align}
    &\ \Re\{\mu(\lambda_{*})\} = 0,
\end{align}
we can write,
\begin{align}
    \frac{d}{d \lambda} \Re\{\mu(\lambda_{*})\} = \frac{d}{d \lambda} \Re \big\{\mu\big(\lambda_0 + k\Delta(\lambda_0)\big)\big\} = 0.
\end{align}
This results in,
\begin{align}
    &\ \mathcal{N}(\lambda_{*})^T\big(I_m + k\Delta_{\lambda_0}^T\big) = 0,
\end{align}
which provides an expression for the sensitivity:
\begin{align}
    &\ \Delta_{\lambda_0} = -\big[k^T\mathcal{N}(\lambda_{*})\big]^{-1} \mathcal{N}(\lambda_{*}).
\end{align}
The parameters $\lambda$ are often classified as $\lambda = (\lambda^{uc}, \lambda^c)$, where $\lambda^{uc}$ represents uncontrollable parameters and $\lambda^{c}$ are the controllable ones. Generally, variations in $\lambda^{uc}$ are beyond the control of operators or designers and may lead to bifurcations, while $\lambda^{c}$ includes design or control parameters that can be maintained or adjusted. Our goal is to modify the controllable parameters $\lambda^{c}$ to enhance robustness against changes in the uncontrollable parameters $\lambda^{uc}$. For this, the normal vector can be partitioned as:
\begin{equation}
    \mathcal{N} = (\mathcal{N}^{uc}, \mathcal{N}^{c}).
\end{equation}
When the parameter variation direction $k = \{(k^{uc}, k^c)\ |\ k^c = 0\}$ has no components for $\lambda^{c}$, we get $k^T\mathcal{N}(\lambda_{*}) = (k^{uc})^T\mathcal{N}^{uc}(\lambda_{*})$. The sensitivity of the margin\footnote{Note that the margin here refers to the margin corresponding to the uncontrollable parameters.} with respect to the controllable parameters $\lambda^c$ is then:
\begin{equation} \label{eqn:s1}
    \Delta_{\lambda^c_0} = -\big[(k^{uc})^T\mathcal{N}^{uc}(\lambda_{*})\big]^{-1} \mathcal{N}^c(\lambda_{*}).
\end{equation}
The sensitivities can be used to determine the optimal controllable parameter in $\lambda^c$ to adjust to increase the margin $\Delta$. 

\section{Modeling} \label{model}
In this section, we describe the detailed model of a droop-controlled GFM inverter connected to an infinite bus. The system setup is depicted in Fig. \ref{fig:setup_gfm}. 
Since our focus is on the system-level dynamics, we make the following assumptions.
\begin{assumption}
    The DC-side dynamics of the source are ignored. That is, the DC voltage $u_{dc}$ remains fixed, while the DC current $i_{dc}$ is an algebraic variable ensuring the required power supply to the inverter.
\end{assumption}
\begin{assumption}
    The PWM switching dynamics are ``fast enough" to be ignored, i.e., the inverter outputs voltage as the computed setpoints $v_{t,d}$ and $v_{t,q}$ instantaneously.
\end{assumption} 
\begin{assumption}
    The transmission line is short and can be modeled using lumped parameters with resistance and inductance in series.
\end{assumption}

\subsection{Reference Frames}
A global rotating $DQ$-frame is established to transform sinusoidal quantities into nearly constant values. The angular velocity (frequency) of this reference frame is defined as $\omega_{DQ} \omega_b$ rad/s, where $\omega_{DQ}$ denotes the per-unit frequency of the $DQ$-frame (often set to be the nominal frequency $\omega_0 = 1$), and $\omega_b$ represents the base frequency, e.g., $2\pi \times 50$ rad/s in Europe and most of Asia, and $2\pi \times 60$ rad/s in North America. 
Voltages and currents throughout the network are expressed in the global $DQ$-frame as,
\begin{equation}
    \begin{aligned}
    &v_t = v_{t,D} + jv_{t,Q}, \\
    &i_t = i_{t,D} + ji_{t,Q}.
\end{aligned}    
\end{equation}
Each inverter operates in a local $dq$-frame rotating at an angular frequency of $(\Delta \omega + \omega_0)\omega_b$ rad/s, where the per-unit frequency deviation $\Delta \omega$ is determined by the inverter's internal control mechanism, such as droop control. Voltages and currents at the inverter terminals are represented in this local $dq$-frame as,
\begin{equation}
    \begin{aligned}
        &v_t = v_{t,d} + jv_{t,q}, \\
        &i_t = i_{t,d} + ji_{t,q}.
    \end{aligned}
\end{equation}
The local $dq$-frame variables need to be transformed into the global $DQ$-frame to establish a uniform system representation. We introduce the rotation matrix $\mathcal{R}(\theta)$,
\begin{equation}
    \mathcal{R}(\theta) = \begin{bmatrix} \cos(\theta) & -\sin(\theta) \\ \sin(\theta) & \cos(\theta) \end{bmatrix},
\end{equation}
where $\theta$ represents the angle of the local $dq$-frame relative to the global $DQ$-frame. The transformation is then given by:
\begin{equation}
    \begin{aligned}
    \begin{bmatrix} v_{t,D} \\ v_{t,Q} \end{bmatrix} &= \mathcal{R}(\theta) \begin{bmatrix} v_{t,d} \\ v_{t,q} \end{bmatrix}, \\
    \begin{bmatrix} i_{t,D} \\ i_{t,Q} \end{bmatrix} &= \mathcal{R}(\theta) \begin{bmatrix} i_{t,d} \\ i_{t,q} \end{bmatrix}.
    \end{aligned}
\end{equation}

\subsection{Grid-Forming Inverter} \label{sec:gfm}
\begin{figure}[ht!]
    \hspace{-2ex} \centerline{\includegraphics[scale=0.7]{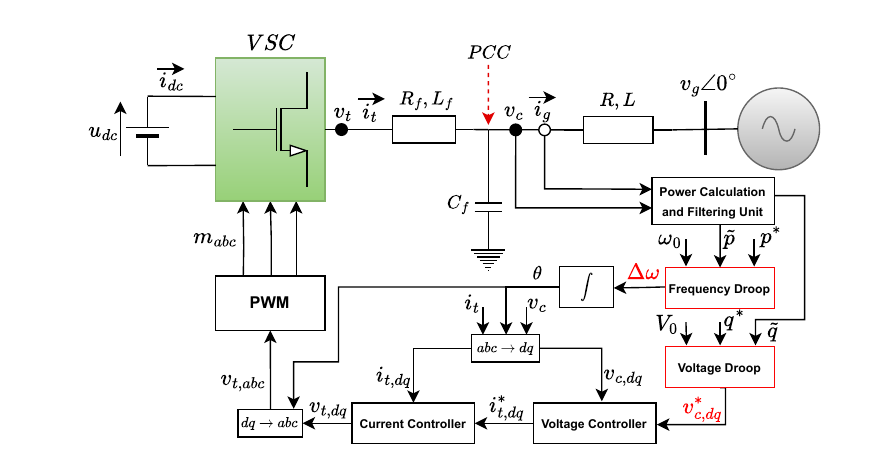}}
    \caption{ GFM inverter connected to the grid.}
    \label{fig:setup_gfm} 
\end{figure}

\begin{figure*}[ht!]
    \centerline{\includegraphics[scale=0.52]{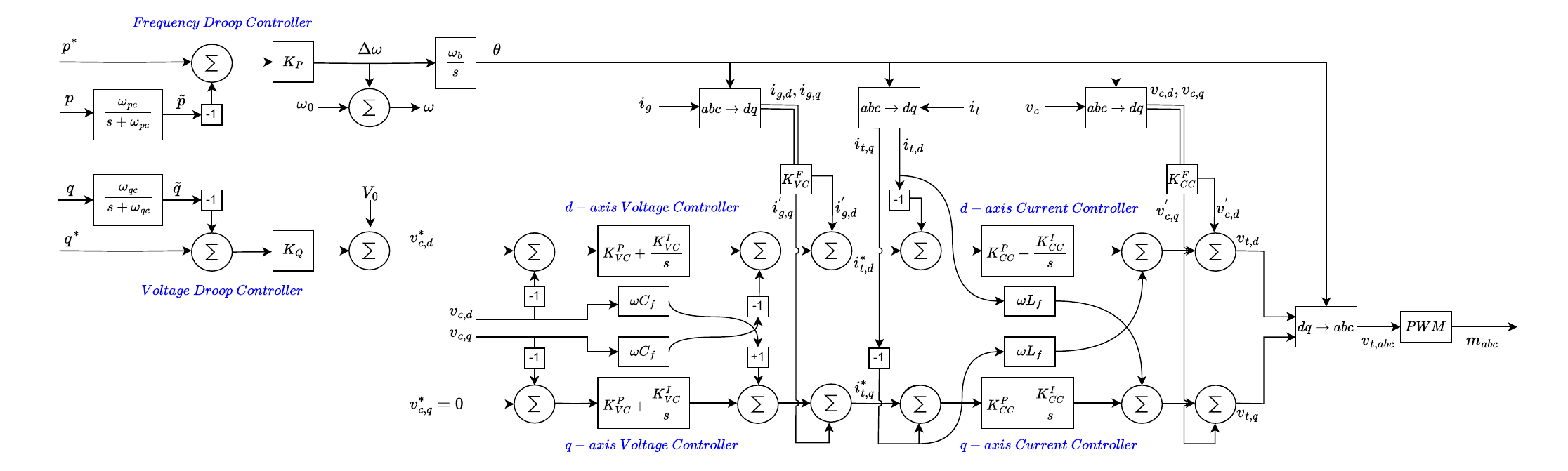}}
    \caption{Detailed control architecture of the droop-based GFM inverter.}
    \label{fig:control_gfm} 
\end{figure*}

A GFM inverter operates as a controllable voltage source behind an output filter, similar to grid-connected synchronous generators. During contingencies, GFM inverters can promptly adjust their output power to balance loads and stabilize local voltage and frequency. To provide context for the analysis, we present a commonly used droop-based GFM control structure. The architecture of a three-phase GFM inverter system is shown in Fig. \ref{fig:setup_gfm}, featuring a PWM-based voltage source converter (VSC) connected with an output {RLC} filter, as described by \eqref{eqn:sm8}-\eqref{eqn:sm11}.
As usual, the grid is modeled as a rigid voltage source connected in series with a transmission line.
The controller has a nested structure with an outer voltage loop and an inner current loop, as given in Fig. \ref{fig:control_gfm}. 
Each control loop is implemented in a local $dq$-frame using Park’s transformation as observed in \eqref{eqn:am7}-\eqref{eqn:am12}. The outer loop comprises the voltage control ({VC}) as described by \eqref{eqn:sm4}-\eqref{eqn:sm5}. The reference signals produced by the outer controller loops viz. \eqref{eqn:am3}-\eqref{eqn:am4}, are sent to the inner current controller ({CC}) as given by \eqref{eqn:sm6}-\eqref{eqn:sm7} and \eqref{eqn:am5}-\eqref{eqn:am6}. The DC power balance equation is given in \eqref{eqn:am15}. GFM control with droop characteristics enables direct voltage and frequency control. The frequency droop mechanism employed in this architecture generates the angle \eqref{eqn:sm3}, \eqref{eqn:am1.5} and frequency \eqref{eqn:am1} setpoints while the voltage droop mechanism generates voltage setpoints as given in \eqref{eqn:am2}. These are supplied with filtered power signals \eqref{eqn:sm1}-\eqref{eqn:sm2} that are derived from the actual values \eqref{eqn:am13}-\eqref{eqn:am14}. The notations are explained in the Appendix~A. 
\subsubsection{Dynamic Equations:}
\begin{align}
       \dot{\tilde{p}} &\ =\ -\omega_{pc} \tilde{p} + p \omega_{pc} \label{eqn:sm1}
        \\
        \dot{\tilde{q}} &\ =\ -\omega_{qc} \tilde{q} + q \omega_{qc} \label{eqn:sm2}
        \\
        \dot{\theta} &\ =\ \omega_b \Delta\omega  \label{eqn:sm3}
        \\
        \dot{\beta_d} &\ =\ v^{*}_{c,d} - v_{c,d} \label{eqn:sm4}
        \\
        \dot{\beta_q} &\ =\ v^{*}_{c,q} - v_{c,q} \label{eqn:sm5}
        \\
         \dot{\gamma_d} &\ =\ i^{*}_{t,d} - i_{t,d} \label{eqn:sm6}
        \\
         \dot{\gamma_q} &\ =\ i^{*}_{t,q} - i_{t,q} \label{eqn:sm7}
         \\
        \dot{v_{c,D}} &\ =\ \omega v_{c,Q} + \frac{1}{C_f}(i_{t,d} - i_{g,d}) \label{eqn:sm8}
        \\
         \dot{v_{c,Q}} &\ =\ -\omega v_{c,D} + \frac{1}{C_f}(i_{t,q} - i_{g,q}) \label{eqn:sm9}
        \\
        \dot{i_{t,d}} &\ =\ \omega i_{t,q} + \frac{1}{L_f}(v_{t,d} - v_{c,d}) - \frac{R_f}{L_f}i_{c,d} \label{eqn:sm10}
        \\
        \dot{i_{t,q}} &\ =\ -\omega i_{t,d} + \frac{1}{L_f}(v_{t,q} - v_{c,q}) - \frac{R_f}{L_f}i_{c,q} \label{eqn:sm11}
\end{align}
\subsubsection{Algebraic Equations:}
\begin{align}
        &\ \omega - \omega_0 - \Delta \omega = 0 \label{eqn:am1}
        \\&\
        \Delta \omega - K_P(p^{*}-\tilde{p}) = 0 \label{eqn:am1.5}
        \\&\
        v^{*}_{c,d} - V_0 - K_Q(q^{*} - \tilde{q}) = 0 \label{eqn:am2}
        \\&\
        i^{*}_{t,d} - K^F_{VC}i_{g,d} \!-\! K^P_{VC}(v^{*}_{c,d} \!-\! v_{c,d}) - K^I_{VC}\beta_d + v_{c,q} \omega C_f \!=\! 0 \label{eqn:am3}
        \\&\
        i^{*}_{t,q} - K^F_{VC}i_{g,q} \!-\! K^P_{VC}(v^{*}_{c,q} \!-\! v_{c,q}) - K^I_{VC}\beta_q - v_{c,d} \omega C_f \!=\! 0 \label{eqn:am4}
        \\&\ v_{t,d} - K^F_{CC}v_{c,d} \!-\! K^P_{CC}(i^{*}_{t,d} \!-\! i_{t,d}) - K^I_{CC}\gamma_d + i_{t,q} \omega L_f \!=\! 0 \label{eqn:am5}
        \\&\
        v_{t,q} - K^F_{CC}v_{c,q} \!-\! K^P_{CC}(i^{*}_{t,q} \!-\! i_{t,q}) - K^I_{CC}\gamma_q - i_{t,d} \omega L_f \!=\! 0 \label{eqn:am6}
        \\&\ 
        i_{g,d} - i_{g,D}\cos(\theta) - i_{g,Q}\sin(\theta) = 0 \label{eqn:am7}
        \\&\
        i_{g,q} + i_{g,D}\sin(\theta) - i_{g,Q}\cos(\theta) = 0 \label{eqn:am8}
        \\&\ 
        i_{t,d} - i_{t,D}\cos(\theta) - i_{t,Q}\sin(\theta) = 0 \label{eqn:am9}
        \\&\
        i_{t,q} + i_{t,D}\sin(\theta) - i_{t,Q}\cos(\theta) = 0 \label{eqn:am10}
        \\&\ v_{c,d} - v_{c,D}\cos(\theta) - v_{c,Q}\sin(\theta) = 0 \label{eqn:am11}
        \\&\
        v_{c,q} + v_{c,D}\sin(\theta) - v_{c,Q}\cos(\theta) = 0 \label{eqn:am12} 
        \\&\
        p - v_{c,d}i_{g,d} - v_{c,q}i_{g,q} = 0 \label{eqn:am13}
        \\&\
        q - v_{c,q}i_{g,d} + v_{c,d}i_{g,q} = 0 \label{eqn:am14}
        \\&\
        u_{dc}i_{dc} - v_{t,d}i_{t,d} - v_{t,q}i_{t,q} = 0 \label{eqn:am15}
\end{align}

\subsection{Dynamic Line Model}
Static phasor-based models are typically adopted for transmission lines in conventional power systems that are dominated by synchronous machines, neglecting transient line dynamics. 
It is also common to assume the line runs at the nominal frequency of $\omega_0$. However, with the increasing penetration of IBRs, incorporating fast line dynamics may be important to capture potential interactions of inverters with the network. Consider a line with a series resistance $R$ and inductance $L$ connecting the inverter to the grid. We can write the dynamic line equations, in per unit, as,
\begin{align}
        &\ \dot{i_{g,D}} - \frac{\omega_{ss}}{L}(v_{c,D}-v_{g,D}) + \frac{R}{L}\omega_{ss}i_{g,D} - \omega_0\omega_{ss}i_{g,Q} = 0,
        \label{eqn:sm12}
        \\&\
        \dot{i_{g,Q}} - \frac{\omega_{ss}}{L}(v_{c,Q}-v_{g,Q}) + \frac{R}{L}\omega_{ss}i_{g,Q} + \omega_0\omega_{ss}i_{g,D} = 0,
        \label{eqn:sm13}
\end{align}
where $\omega_{ss}$ is the frequency of the system at steady state. When an infinite bus exists in the system, $\omega_{ss}=\omega_0$.

\subsection{Static Line Model}
The electrical variables observed from the global $DQ$-frame become constant at steady state if the choice of $\omega_{DQ}$ matches the steady-state frequency. Hence, the dynamic line model \eqref{eqn:sm12}-\eqref{eqn:sm13} reverts to a set of algebraic equations at steady-state as given by, 
\begin{align}
        &\ i_{g,D} - \frac{R}{Z^2}(v_{c,D} - v_{g,D}) - \frac{X}{Z^2}(v_{c,Q} - v_{g,Q}) = 0, \label{eqn:am16}
        \\&\
        i_{g,Q} - \frac{R}{Z^2}(v_{c,Q} - v_{g,Q}) + \frac{X}{Z^2}(v_{c,D} - v_{g,D}) = 0, \label{eqn:am17}
        \\&\
        Z^2 - R^2 - X^2 = 0. \label{eqn:am18}
\end{align}

Power system models are generally represented by {differential-algebraic equations} (DAEs).
However, one of the challenges with the normal vector method is that it only works directly with ordinary differential equations (ODEs).
However, the main nonlinear complexity of the single-inverter-infinite-bus system lies in one state variable, $\theta$, while the rest are in bilinear form. By explicitly solving the algebraic equations, we can convert the DAEs into ODEs analytically. This transformation yields a set of state equations in ODE form, as shown in Appendix~B. This step simplifies the computation of Jacobians and Hessians needed to obtain the normal vector, and it allows us to derive a closed-form expression for sensitivity in \eqref{eqn:s1}.

\section{Results} \label{results}
This section presents the numerical results on parameter sensitivities to Hopf bifurcations and the mechanism of tuning controllable parameters to mitigate oscillatory instability for GFM IBRs. We also discuss the results on the impacts of line dynamics on stability margin. 

\subsection{System Setup and Stability Margin to Hopf Bifurcation}
The system is shown in Fig.~\ref{fig:setup_gfm}. All variables and parameters are converted to per-unit quantities using base values of 1052KVA and 320KV. The system includes controllable and uncontrollable parameters, identified as follows, 
\begin{equation*}
    \begin{aligned}
        \lambda^{uc} & := \big(R_f,L_f,C_f,\omega_{b,ss},\omega_{pc,qc},v^{*}_{c,q},v_{g,\{D,Q\}},X,R\big) \in \mathbb{R}^{12},
        \\
        \lambda^{c} & := \big(K^{P,I,F}_{VC},K^{P,I,F}_{CC},K_{P,Q},\omega_0,V_0,p^{*},q^{*}\big) \in \mathbb{R}^{12}.
    \end{aligned}
\end{equation*} 
We aim to identify the most impactful controllable parameters to counteract the effects of variations in uncontrollable parameters. 

Table \ref{tab:GFM_param} shows the nominal parameter values and specifies the value at which each parameter triggers a Hopf bifurcation when varied individually. Parameters not analyzed for bifurcations due to engineering considerations are marked with a “-,” while those that do not cause any Hopf bifurcation are indicated with a “\xmark.” 

The table also provides results for systems with static and dynamic transmission line models. Upon examining the last two columns, an important observation is made: including transmission line dynamics in the model reduces the stability margin across all parameters in this system. This suggests that neglecting transmission line dynamics could lead to optimistic conclusions regarding stability. However, as shown in the table, the differences are generally small for most parameters, with one exception of the voltage-reactive power droop gain ($K_Q$), where the gap is more significant.

\begin{table*}[htb]
  \setlength\tabcolsep{0pt}
  \setlength\extrarowheight{2pt}
  \caption{Parameter values and stability margins in the systems with static and dynamic lines.}
  \makebox[\textwidth][c]{
    \begin{tabular*}{\textwidth}{@{\extracolsep{\fill}}*{6}{c}}
      \Xhline{2\arrayrulewidth} 
      \multirow{2}{*}{Parameter} & \multirow{2}{*}{Nominal value} & \multicolumn{2}{c}{Value at Hopf Bifurcation} & \multicolumn{2}{c}{Margin} \\
      \cline{3-4} \cline{5-6} 
       & & \multicolumn{1}{c}{S-line} & \multicolumn{1}{c}{D-line} & \multicolumn{1}{c}{S-line} & \multicolumn{1}{c}{D-line} \\
      \hline
      $R_f$ & 0.0072 p.u. & 1.65282 p.u. & 1.65073 p.u. & 1.64562 p.u. & 1.64353 p.u.  \\
      $L_f$ & 0.05 p.u. & 0.126864 p.u. & 0.123537 p.u. & 0.076864 p.u. & 0.073537 p.u.  \\
      $C_f$ & 0.3 p.u. & 2.20075 p.u. & 2.18145 p.u. & 1.90075 p.u. & 1.88145 p.u.  \\
      \hline 
      $\omega_{pc}$ & 332.8 rad/s & 11.5235 rad/s & 12.016 rad/s & 321.2765 rad/s & 320.784 rad/s  \\
      $\omega_{qc}$ & 732.8 rad/s & \xmark & \xmark & \xmark & \xmark  \\
      \hline
      $K_P$ & 1.8 \% & 6.4585 \% & 6.0154 \% & 4.6585 \% & 4.2154 \%  \\
      $K_Q$ & 0.01 \% & 39.418 \% & 23.72405 \% & 39.408 \% & 23.71405 \%  \\
      \hline
      $K^P_{VC}$ & 1 p.u. & 0.14505 p.u. & 0.1532 p.u. & 0.85495 p.u. & 0.8468 p.u.  \\
      $K^I_{VC}$ & 1.16 p.u. & 6.2701 p.u. & 6.013 p.u. & 5.1101 p.u. & 4.853 p.u.  \\
      $K^F_{VC}$ & 1 p.u. & 1.263435 p.u. & 1.24693 p.u. & 0.263435 p.u. & 0.24693 p.u.  \\
      \hline
      $K^P_{CC}$ & 2.5 p.u. & 0.82689 p.u. & 0.85645 p.u. & 1.67311 p.u. & 1.64355 p.u.  \\
      $K^I_{CC}$ & 1.19 p.u. & \xmark & \xmark & \xmark & \xmark \\
      $K^F_{CC}$ & 0 p.u. & 1.9071 p.u. & 1.8842 p.u. & 1.9071 p.u. & 1.8842 p.u.  \\
      \hline
      $\omega_0$ & 1 p.u. & - & - & - & -  \\
      $V_0$ & 1 p.u. & - & - & - & -  \\
      $p^{*}$ & 1 p.u. & \xmark & \xmark & \xmark & \xmark  \\
      $q^{*}$ & 0.5 p.u. & \xmark & \xmark & \xmark & \xmark  \\
      \hline
      $X$ & 0.2 p.u. & 0.08404 p.u. & 0.08843 p.u. & 0.11596 p.u. & 0.11157 p.u.  \\
      $R$ & 0.02 p.u. & \xmark & \xmark & \xmark & \xmark  \\
      \Xhline{2\arrayrulewidth}
     \end{tabular*}
     }
    \label{tab:GFM_param}
\end{table*}

\subsection{Estimation of Stability Margin Using Normal Vectors}
The onset of the Hopf bifurcation can be significantly delayed by using sensitivity information derived from the normal vector method. 

Let’s consider a case where the parameter $I$ changes thereby leading to instability, and we use parameter $C$ as the control parameter to counteract this destabilizing effect. Once these two parameters and the direction of change in parameter $I$ are selected, the sensitivity formula in \eqref{eqn:s1} allows us to directly calculate the estimated stability margin $\hat{\Delta}^I_\text{new}$ based on changes in the control parameter,

\begin{equation} \label{eqn:pred}
    \hat{\Delta}^I_\text{new} = \Delta^I_\text{old} + \Delta^{C|I}_{\lambda_0}\big(\lambda^C_\text{new} - \lambda^C_\text{old}\big),
\end{equation}
where $I$ and $C$ represent the indices of the instability-causing and control parameters, respectively. On the other hand, the actual margin is numerically found as follows,
\begin{align} \label{eqn:true}
   \Delta^I_\text{new} &= \Delta^I_\text{old} + \big(\lambda^I_{*,\text{new}} - \lambda^I_{*,\text{old}}\big) \nonumber \\
   & = \lambda^I_{*,\text{new}} - \lambda^I_{0}.
\end{align}

We use an example to demonstrate the normal vectors approach in obtaining an efficient estimation of the stability margin. Consider a scenario where the grid impedance $X$ decreases, indicating stronger grid conditions, and we use $K^F_{VC}$ as the control parameter to mitigate this destabilizing effect. After selecting these two parameters and the direction of $X$ increasing, the sensitivity formula in \eqref{eqn:pred} enables us to compute the estimated stability margin based on the adjustments in $K^F_{VC}$.  

\begin{table}[ht!]
\begin{center}
\caption{Effects of parameter variation on Hopf stability margin\\ (static line)}
\begin{tabular}{cccc}
\Xhline{2\arrayrulewidth}
\multirow{2}{*}{$K^F_{VC}\ \mac{\downarrow}$} & Estimated Margin & True Margin & \multirow{2}{*}{Error} \\
& in X $\mac{\uparrow}$ & in X & \\
\hline
        1 (Nominal) & - & 0.11596 & - \\
        0.98 & 0.13395 & 0.13538 & 1.43e-03 \\
         0.95 & 0.16093 & 0.16613 & 5.203e-03 \\
         0.92 & 0.18791 & 0.18666 & -1.25e-03 \\
          \Xhline{2\arrayrulewidth}
\end{tabular}
\label{tab:GFM_H_sim_stat}
\end{center}
\end{table}

Table \ref{tab:GFM_H_sim_stat} compares the true margin with the estimated margin given by the normal vector method for the system with a static transmission line, while Table \ref{tab:GFM_H_sim_dyn} does the same for the case with a dynamic line. 
The first rows in Tables \ref{tab:GFM_H_sim_stat} and \ref{tab:GFM_H_sim_dyn} show the nominal values of $K^F_{VC}$ and the (nominal) true margin in $X$. Subsequent rows display reductions in $K^F_{VC}$ and the corresponding increases in the stability margin of $X$ for both the estimated and true values.
It can be observed from both tables that the normal vector method effectively estimates the sensitivity margins. 

\begin{table}[ht!]
\begin{center}
\caption{Effects of parameter variation on Hopf stability margin\\ (dynamic line)}
\begin{tabular}{cccc}
\Xhline{2\arrayrulewidth}
\multirow{2}{*}{$K^F_{VC}\ \mac{\downarrow}$} & Estimated Margin & True Margin & \multirow{2}{*}{Error} \\
& in X $\mac{\uparrow}$ & in X & \\
\hline
        1 (Nominal) & - & 0.11157 & - \\
         0.98 & 0.12981 & 0.13164 & 1.83e-03 \\
         0.95  & 0.15717 & 0.16486 & 7.69e-03 \\
         0.92  & 0.18453 & 0.18654 & 2.01e-03 \\
          \Xhline{2\arrayrulewidth}
\end{tabular}
\label{tab:GFM_H_sim_dyn}
\end{center}
\end{table}

The theoretical results on parameter sensitivities are validated using time-domain simulations. 
Referring to Table \ref{tab:GFM_param}, $X$ is reduced to $0.08404$ p.u. and $0.08843$ p.u., respectively, for the static and dynamic line scenarios. This leads to Hopf bifurcations in the respective cases with the nominal value of $K^F_{VC} = 1$.   
\begin{figure}[ht!]
    \hspace{-1ex} \centerline{\includegraphics[scale=0.36]{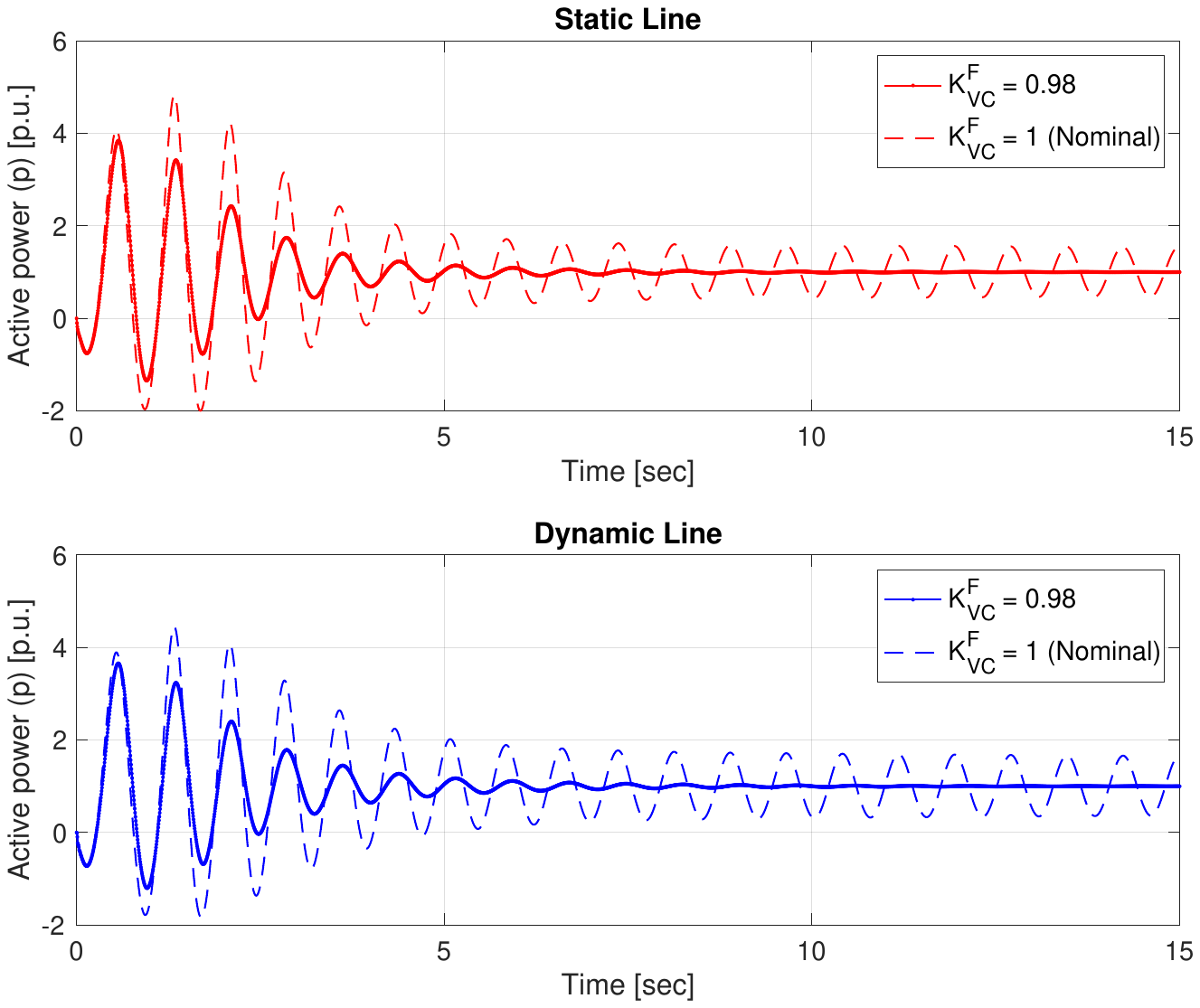}}
    \caption{Trajectories of the active power output ($p$) under various parameter values of $K^F_{VC}$ with static (upper figure) and dynamic (lower figure) lines. The dashed lines correspond to Hopf bifurcation, while the solid lines correspond to stable equilibria.}
    \label{fig:gfm_sim_p} 
\end{figure}
\begin{figure}[ht!]
    \hspace{-1ex} \centerline{\includegraphics[scale=0.36]{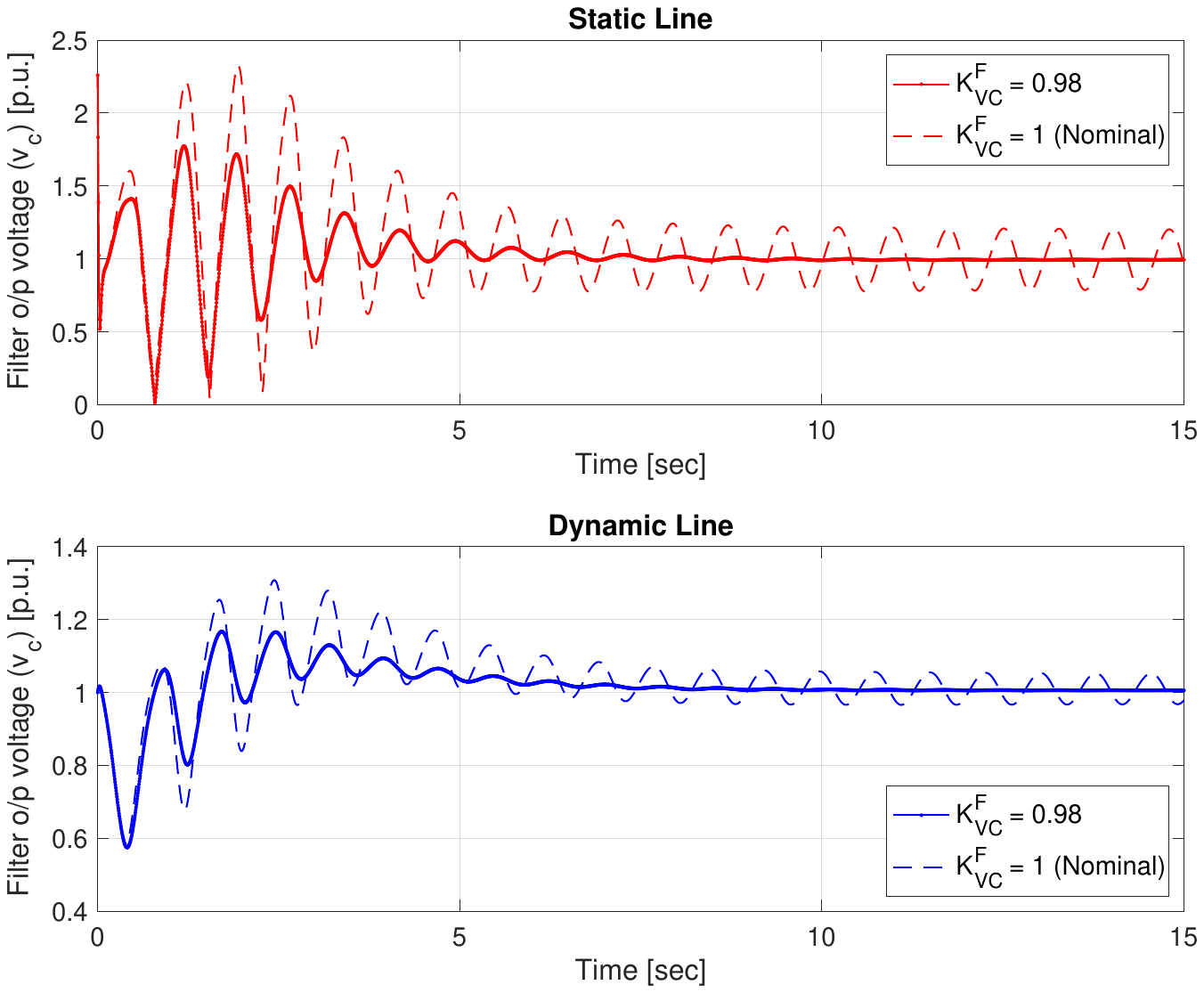}}
    \caption{Trajectories of the filter output voltage ($v_c$) under various parameter values of $K^F_{VC}$ with static (upper figure) and dynamic (lower figure) lines. The dashed lines correspond to Hopf bifurcation, while the solid lines correspond to stable equilibria.}
    \label{fig:gfm_sim_v} 
\end{figure}
As shown in Figs. \ref{fig:gfm_sim_p} and \ref{fig:gfm_sim_v}, when $K^F_{VC}$ is changed to $0.98$ p.u. (corresponding to the second rows of Tables~\ref{tab:GFM_H_sim_stat} and \ref{tab:GFM_H_sim_dyn}), the oscillations cease to exist in both cases. This corresponds to positive stability margins exhibited in Tables \ref{tab:GFM_H_sim_stat} and \ref{tab:GFM_H_sim_dyn}.

\subsection{Complete Analysis of Control Parameter Sensitivity}
\begin{figure}[ht!]
    \hspace{8ex}\centerline{\includegraphics[scale=0.36]{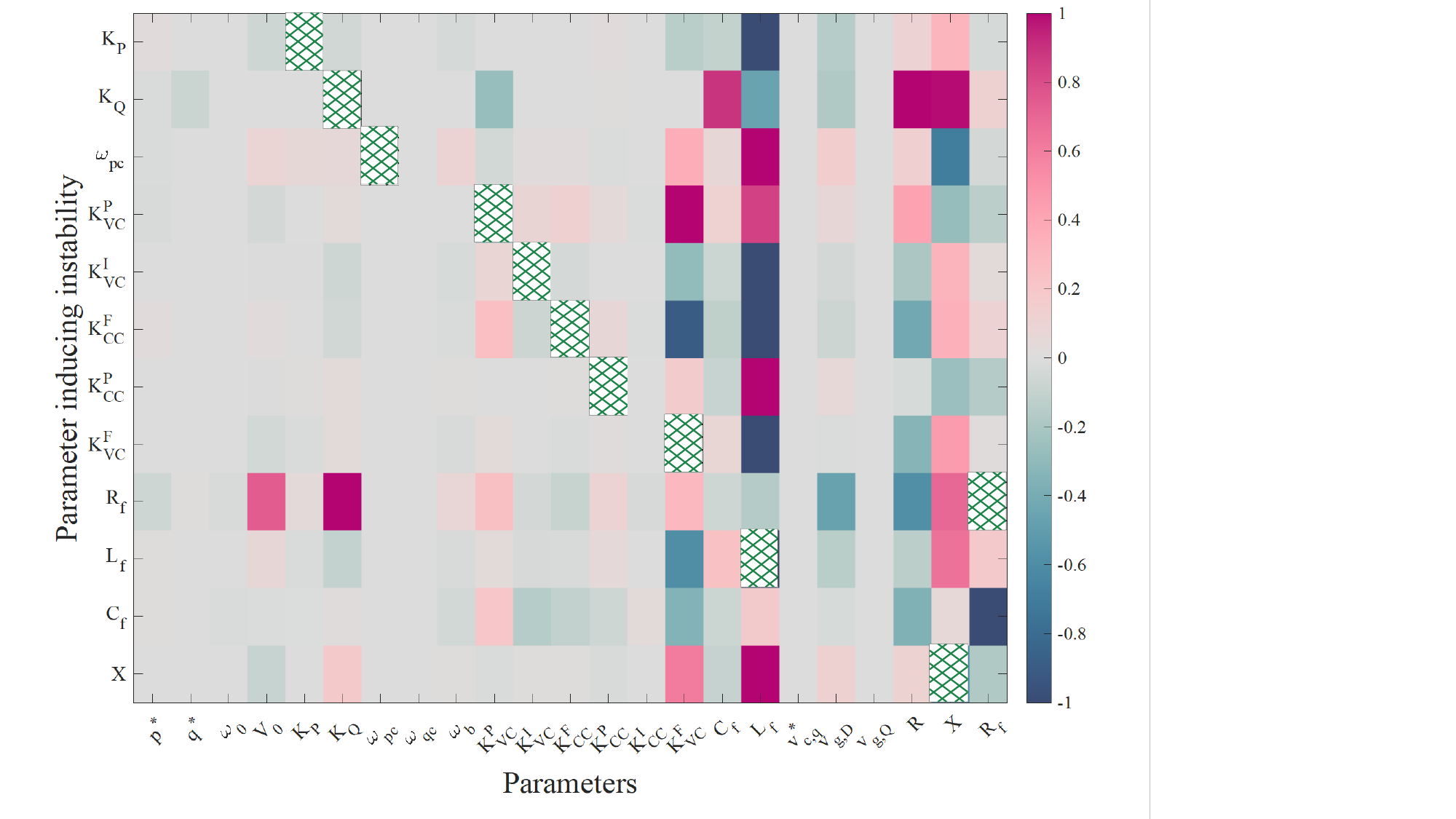}}
    \caption{Heatmap of parameter sensitivities to Hopf bifurcation for the system where GFM is connected to the grid via a static line. The heatmap is normalized using maximum-absolute row scaling. Green criss-cross entries have the same x and y indices and are therefore meaningless.}
    \label{fig:heatmap_GFM_H_stat} 
\end{figure}

\begin{figure}[ht!]
    \hspace{8ex}\centerline{\includegraphics[scale=0.36]{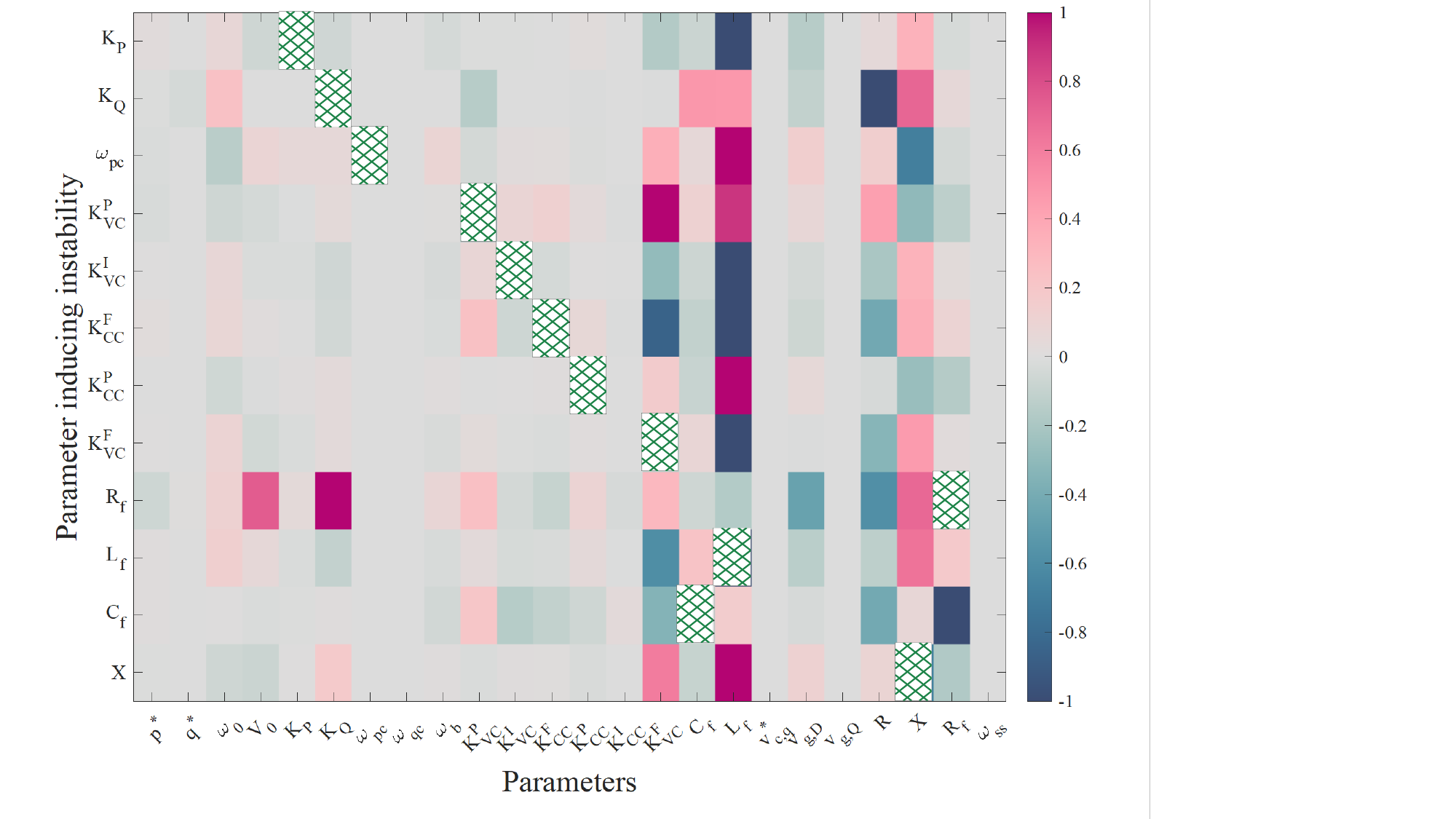}}
    \caption{Heatmap of parameter sensitivities to Hopf bifurcation for the system where GFM is connected to the grid via a dynamic line. The heatmap is normalized using maximum-absolute row scaling. Green criss-cross entries have the same x and y indices and are therefore meaningless.}
    \label{fig:heatmap_GFM_H_dyn} 
\end{figure}

In Figs. \ref{fig:heatmap_GFM_H_stat} and \ref{fig:heatmap_GFM_H_dyn}, we generate normalized heatmaps for the parameter sensitivities to Hopf bifurcation with static and dynamic transmission lines, respectively.
The parameters shown on the y-axis of the map are responsible for triggering Hopf bifurcation, while those on the x-axis represent the parameters that can be adjusted to improve the stability margin\footnote{For completeness, we include all parameters on the x-axis despite that only part of them are controllable in practice.}. For instance, according to the map, if instability occurs due to a reduction in $K^P_{VC}$, the most effective adjustment would be to decrease $K^F_{VC}$, as it displays the darkest shade (red, in this case) among all the controllable parameters. The green criss-cross entries with the same x and y indices are meaningless, as they are always -1. In other words, if the parameter causing instability is known, the best solution is to adjust that parameter in the opposite direction. 
Additionally, positive (or negative) sensitivity values for control parameters do not automatically imply that increasing (or decreasing) their values will improve the margin. The effect also depends on the direction of change in the instability-causing parameter.

Table \ref{tab:GFM_H_infl} summarizes the most influential parameters that can help mitigate instabilities caused by \textit{other} parameters. The colored up and down arrows indicate the direction in which parameter changes lead to instabilities. The \xmark\ symbol signifies the absence of a Hopf bifurcation. The sensitivity $\Delta^{C|I}_{\lambda_0}$ is expressed in per unit.
\begin{table*}[ht!]
  \begin{center}
  \setlength\tabcolsep{6pt}
  \setlength\extrarowheight{2pt}
    \caption{Most Influential Control Parameters for GFM Inverter}
    \begin{tabular}{cccccc}
        \Xhline{2\arrayrulewidth}
        \multirow{2}{*}{Cause of} & \multicolumn{2}{c}{Static Line} & \multicolumn{2}{c}{Dynamic Line} \\
        \cmidrule(lr){2-3}
        \cmidrule(lr){4-5} 
        instability (I) & Control (C) & Sensitivity \big($\Delta^{C|I}_{\lambda_0}$\big) & Control (C) & Sensitivity \big($\Delta^{C|I}_{\lambda_0}$\big)  \\
        \hline
        $K_P\ \mac{\uparrow}$ & $\cac{K^F_{VC}}$ & -11.5938 & $\cac{K^F_{VC}}$ & -12.7678 \\
        $K_Q\ \mac{\uparrow}$ & $K^P_{VC}$ & -38.4021 & $K^P_{VC}$ & -23.23 \\
        $\omega_{pc}\ \mac{\downarrow}$ & $\cac{K^F_{VC}}$ & 70.7682 & $\cac{K^F_{VC}}$ & 75.6362 \\
        $K^P_{VC}\ \mac{\downarrow}$ & $\cac{K^F_{VC}}$ & 3.5189 & $\cac{K^F_{VC}}$ & 3.5794 \\
        $K^I_{VC}\ \mac{\uparrow}$ & $\cac{K^F_{VC}}$ & -31.8407 & $\cac{K^F_{VC}}$ & -30.7589 \\
        $K^F_{VC}\ \mac{\uparrow}$ & $K_Q$ & 0.3041 & $K_Q$ & 0.2790 \\
        $K^P_{CC}\ \mac{\downarrow}$ & $\cac{K^F_{VC}}$ & 3.2088 & $\cac{K^F_{VC}}$ & 3.4463\\
        $K^I_{CC}$ & \xmark & \xmark & \xmark & \xmark \\
        $K^F_{CC}\ \mac{\uparrow}$ & $\cac{K^F_{VC}}$ & -7.9018 & $\cac{K^F_{VC}}$ & -8.0179 \\
        $R_f\ \mac{\uparrow}$ & $K_Q$ & 5.6254 & $K_Q$ & 5.6326\\
        $L_f\ \mac{\uparrow}$ & $\cac{K^F_{VC}}$ & -0.5887 & $\cac{K^F_{VC}}$ & -0.5904\\
        $C_f\ \mac{\uparrow}$ & $\cac{K^F_{VC}}$ & -4.9578 & $\cac{K^F_{VC}}$ & -4.9364\\
        $X\ \mac{\downarrow}$ & $\cac{K^F_{VC}}$ & 0.8994 & $\cac{K^F_{VC}}$ & 0.912 \\
        $R$ & \xmark & \xmark & \xmark & \xmark \\
        \Xhline{2\arrayrulewidth}
    \end{tabular}
    \label{tab:GFM_H_infl}
    \end{center}
\end{table*}
Table \ref{tab:GFM_H_infl} shows that the integral gain of the current controller $K^I_{CC}$ and transmission line resistance $R$ do not induce Hopf bifurcation in this specific GFM inverter-based system.
Most importantly, $K^F_{VC}$ turns out to be the most effective control parameter for preventing oscillatory instability caused by variations in most other parameters. The only exceptions are changes in the filter resistance and reactive power droop, where adjusting the $K_Q$ and $K^P_{VC}$ respectively, prove to be the most effective control actions. 
To draw a comparison, the author's prior work \cite{chatterjee2024sensitivity} identified a consistent control parameter for the grid-following inverter system—the proportional gain of the current control loop—that could help steer the system away from Hopf bifurcations regardless of which parameter is causing instability.

\section{Conclusion}
This paper studied the oscillatory instability of grid-forming (GFM) inverters through the lens of Hopf bifurcation. Using an analytical expression for the sensitivity of the stability margin based on normal vector methods, we conducted a comprehensive investigation of the entire parameter space of the GFM inverter-based system. 
The analysis identified an effective control parameter, i.e., the feedforward gain of the voltage control loop, in counteracting the instability-inducing effect of most of the other parameters.
Notably, the investigation revealed that by incorporating line dynamics, there is a systematic reduction in the stability margin to Hopf bifurcation compared to the commonly used static line models. Nevertheless, such reductions are relatively small. Depending on the problem at hand, engineers can determine whether or not to adopt the dynamic line model. 
The normal vector method is shown to be computationally efficient in finding the stability margin. It gives complete information on the parameter sensitivity once the normal vector is constructed. 
Future work will extend the method and analysis to large networks. In this context, the choice between dynamic and static line models will be critical, given the significant difference in system order and the resulting simulation complexity when modeling large networks with tens of thousands of transmission lines.

\section*{Appendix}
\subsection{Nomenclature} \label{app:a}
The indexing of the variables and parameters used in the differential and algebraic equations is described below.\\
\texttt{Dynamic states (11):}\\
\begin{small}
$x_1\ :\ \Delta \omega \rightarrow$ Angular frequency error\\
$x_2\ :\ \Delta q \rightarrow$ Reactive power measured output\\
$x_3\ :\ \theta \rightarrow$ Load angle\\
$x_4\ :\ \beta_d \rightarrow$ d-axis voltage controller state\\
$x_5\ :\ \beta_q \rightarrow$ q-axis voltage controller state\\
$x_6\ :\ \gamma_d \rightarrow$ d-axis current controller state\\
$x_7\ :\ \gamma_q \rightarrow$ q-axis current controller state\\
$x_8\ :\ v_{c,D} \rightarrow$ (Global) D-axis filter output voltage $v_c$\\
$x_9\ :\ v_{c,Q} \rightarrow$ (Global) Q-axis filter output voltage $v_c$\\
$x_{10}\ :\ i_{t,d} \rightarrow$ (Local)  d-axis IBR output current $i_t$\\
$x_{11}\ :\ i_{t,q} \rightarrow$ (Local) q-axis IBR output current $i_t$
\end{small}
\newline\newline
\texttt{Algebraic states (19):}\\
\begin{small}
$y_1\ :\ \omega \rightarrow$ Angular frequency\\
$y_2\ :\ v^{*}_{c,d} \rightarrow$ d-axis voltage setpoint $v^{*}_c$\\
$y_3\ :\ v^{*}_{c,q} \rightarrow$ q-axis voltage setpoint $v^{*}_c$\\
$y_4\ :\ i_{t,d}^{*} \rightarrow$ d-axis current setpoint $i_t^{*}$\\
$y_5\ :\ i_{t,q}^{*} \rightarrow$ q-axis current setpoint $i_t^{*}$\\
$y_6\ :\ i_{g,d} \rightarrow$ d-axis filter output current $i_g$\\
$y_7\ :\ i_{g,q} \rightarrow$ q-axis filter output current $i_g$\\
$y_8\ :\ i_{g,D} \rightarrow$ D-axis filter output current $i_g$\\
$y_9\ :\ i_{g,Q} \rightarrow$ Q-axis filter output current $i_g$\\
$y_{10}\ :\ i_{t,D} \rightarrow$ D-axis IBR output current $i_t$\\
$y_{11}\ :\ i_{t,Q} \rightarrow$ Q-axis IBR output current $i_t$\\
$y_{12}\ :\ v_{c,d} \rightarrow$ d-axis filter output voltage $v_c$\\
$y_{13}\ :\ v_{c,q} \rightarrow$ q-axis filter output voltage $v_c$\\
$y_{14}\ :\ p \rightarrow$ Active power output from filter\\
$y_{15}\ :\ q \rightarrow$ Reactive power output from filter\\
$y_{16}\ :\ v_{t,d} \rightarrow$ d-axis IBR output voltage $v_t$\\
$y_{17}\ :\ v_{t,q} \rightarrow$ q-axis IBR output voltage $v_t$\\
$y_{18}\ :\ u_{dc} \rightarrow$ DC-link voltage\\
$y_{19}\ :\ i_{dc} \rightarrow$ DC-link current
\end{small}
\newline\newline
\texttt{Parameters (22):}\\
\begin{small}
$p_1\ :\ p^{*} \rightarrow$ Active power setpoint \\
$p_2\ :\ q^{*} \rightarrow$ Reactive power setpoint \\
$p_3\ :\ \omega_0 \rightarrow$ Angular frequency setpoint of IBR \\
$p_4\ :\ V_0 \rightarrow$ Voltage setpoint of IBR \\
$p_5\ :\ K_P \rightarrow$ Active droop coefficient \\
$p_6\ :\ K_Q \rightarrow$ Reactive droop coefficient \\
$p_7\ :\ \omega_{pc} \rightarrow$  Active power filter 3dB cut-off frequency \\
$p_8\ :\ \omega_{qc} \rightarrow$  Reactive power filter 3dB cut-off frequency \\
$p_9\ :\ \omega_b \rightarrow$ Base angular frequency \\
$p_{10}\ :\ K^P_{VC} \rightarrow$  Proportional gain of voltage controller \\
$p_{11}\ :\ K^I_{VC} \rightarrow$  Integral gain of voltage controller \\
$p_{12}\ :\ K^F_{CC} \rightarrow$  Feed-forward gain of voltage controller \\
$p_{13}\ :\ K^P_{CC} \rightarrow$  Proportional gain of current controller \\
$p_{14}\ :\ K^I_{CC} \rightarrow$  Integral gain of current controller \\
$p_{15}\ :\ K^F_{VC} \rightarrow$  Feed-forward gain of current controller \\
$p_{16}\ :\ R_f \rightarrow$ Filter resistance \\
$p_{17}\ :\ C_f \rightarrow$ Filter capacitance \\
$p_{18}\ :\ L_f \rightarrow$ Filter inductance \\
$p_{19}\ :\ v_{g,D} \rightarrow$ D-axis grid voltage $v_g$ \\
$p_{20}\ :\ v_{g,Q} \rightarrow$ Q-axis grid voltage $v_g$ \\
$p_{21}\ :\ R \rightarrow$ Transmission line resistance \\
$p_{22}\ :\ X \rightarrow$ Transmission line reactance
\end{small}

\subsection{Restructured System Equations} \label{app:b}

We analytically eliminate the algebraic variables from the DAEs in Sec. \eqref{sec:gfm} to obtain the following ODEs for the GFM inverter-based system.
\begin{align*}
        \dot{\tilde{p}} &\ = -\omega_{pc} \tilde{p} + \Bigg(\frac{R}{Z^2}\Big[v_{c,D}^2 + v_{c,Q}^2 - v_{c,D}v_{g,D} \Big]\ +
        \\&\
        \frac{X}{Z^2}v_{c,Q}v_{g,D}\Bigg) \omega_{pc}
        \\
        \dot{\tilde{q}} &\ = -\omega_{qc} \tilde{q} + \Bigg(\frac{X}{Z^2}\Big[v_{c,D}^2 + v_{c,Q}^2 - v_{c,D}v_{g,D} \Big]\ - 
        \\&\
        \frac{R}{Z^2}v_{c,Q}v_{g,D}\Bigg) \omega_{qc}
        \\
        \dot{\theta} &\ = \omega_b \big(p^{*} - \tilde{p}\big) K_P
        \\
        \dot{\beta_d} &\ = V_0 + K_Q(q^{*} - \tilde{q}) - v_{c,D}\cos(\theta) - v_{c,Q}\sin(\theta)
        \\
        \dot{\beta_q} &\ = v_{c,D}\sin(\theta) - v_{c,Q}\cos(\theta)
        \\
        \dot{\gamma_d} &\ = K^F_{VC}\Bigg(\bigg[\frac{R}{Z^2}(v_{c,D} - v_{g,D}) + \frac{X}{Z^2}v_{c,Q} \bigg] \cos(\theta)\ + 
        \\&\
        \bigg[\frac{R}{Z^2}v_{c,Q}\ -\frac{X}{Z^2}(v_{c,D} - v_{g,D}) \bigg] \sin(\theta) \Bigg) + K^I_{VC} \beta_d - i_{t,d}\  
        \\&\
        +K^P_{VC}\Big(V_0 + K_Q\big[q^{*} - \tilde{q}\big] - v_{c,D}\cos(\theta) - v_{c,Q}\sin(\theta) \Big)\ 
         \\&\
         +\Big(v_{c,D}\sin(\theta) - v_{c,Q}\cos(\theta) \Big)\big(\omega_0 + K_P\big[p^{*}-\tilde{p}\big]\big) C_f 
        \\
        \dot{\gamma_q} &\ = K^F_{VC}\Bigg(-\bigg[\frac{R}{Z^2}(v_{c,D} - v_{g,D}) + \frac{X}{Z^2}v_{c,Q} \bigg] \sin(\theta) +  
        \\&\
        \bigg[\frac{R}{Z^2}v_{c,Q} - \frac{X}{Z^2}(v_{c,D} - v_{g,D}) \bigg] \cos(\theta) \Bigg) + K^I_{VC} \beta_q - i_{t,q} 
        \\&\
         + K^P_{VC}\Big(v_{c,D}\sin(\theta) -v_{c,Q}\cos(\theta) \Big) 
         \\&\
         + \Big(v_{c,D}\cos(\theta) + v_{c,Q}\sin(\theta) \Big)\big(\omega_0 + K_P\big[p^{*}-\tilde{p}\big]\big) C_f 
         \\
        \dot{v_{c,D}} &\ = \big(\omega_0 + K_P\big[p^{*}-\tilde{p}\big]\big) v_{c,Q} + \frac{1}{C_f}\Big[i_{t,d}\cos(\theta)\ - 
        \\&\
        i_{t,q}\sin(\theta) - \frac{R}{Z^2} \big(v_{c,D} - v_{g,D}\big) - \frac{X}{Z^2}v_{c,Q}\Big]
        \\
        \dot{v_{c,Q}} &\ = -\big(\omega_0 + K_P\big[p^{*}-\tilde{p}\big]\big) v_{c,D} + \frac{1}{C_f}\Big[i_{t,d}\sin(\theta)\ + 
        \\&\
        i_{t,q}\cos(\theta) - \frac{R}{Z^2}v_{c,Q} + \frac{X}{Z^2}\big(v_{c,D} - v_{g,D}\big)\Big]
        \\
        \dot{i_{t,d}} &\ = \frac{1}{L_f}\Bigg(\big[K^F_{CC}-1\big]\big(v_{c,D}\cos(\theta) + v_{c,Q}\sin(\theta)\big)\ + 
        \\&\
        K^I_{CC}\gamma_d + K^P_{CC}\bigg[\big(v_{c,D}\sin(\theta) - v_{c,Q}\cos(\theta)\big)\big(\omega_0\ +  
        \\&\ 
        K_P\big[p^{*}-\tilde{p}\big]\big)C_f + K^I_{VC}\beta_d - i_{t,d} + K^P_{VC}\big[V_0\ + 
        \\&\
        K_Q\big(q^{*} - \tilde{q}\big) - v_{c,D}\cos(\theta) - v_{c,Q}\sin(\theta)\big] + K^F_{VC}
        \\&\ 
        \Big\{\Big [\frac{R}{Z^2}(v_{c,D} - v_{g,D}) + \frac{X}{Z^2}v_{c,Q} \Big] \cos(\theta)\ + 
        \\&\
        \Big[\frac{R}{Z^2}v_{c,Q} - \frac{X}{Z^2}(v_{c,D} - v_{g,D}) \Big] \sin(\theta) \Big\} \bigg] \Bigg) - \frac{R_f}{L_f}i_{c,d}
        \\
        \dot{i_{t,q}} &\ = \frac{1}{L_f}\Bigg(\big[-K^F_{CC}+1\big]\big(v_{c,D}\sin(\theta) - v_{c,Q}\cos(\theta)\big)\ +  
        \\&\
        K^I_{CC}\gamma_q + K^P_{CC}\bigg[\big(v_{c,D}\cos(\theta) + v_{c,Q}\sin(\theta)\big)\big(\omega_0\ +  
        \\&\ 
        K_P\big[p^{*}-\tilde{p}\big]\big)C_f + K^I_{VC}\beta_q - i_{t,q} + K^P_{VC}\big[v_{c,D}\sin(\theta)\ -
        \\&\
        v_{c,Q}\cos(\theta)\big] + K^F_{VC}\Big\{-\Big[\frac{R}{Z^2}(v_{c,D} - v_{g,D}) + \frac{X}{Z^2}v_{c,Q} \Big]  
        \\&\
        \sin(\theta) + \Big[\frac{R}{Z^2}v_{c,Q} - \frac{X}{Z^2}(v_{c,D} - v_{g,D}) \Big] \cos(\theta) \Big\} \bigg] \Bigg)\ -
        \\&\
        \frac{R_f}{L_f}i_{c,q}
\end{align*}

\bibliographystyle{ieeetr}
\bibliography{Bibliography.bib}
\balance
\end{document}